\documentclass[aps,prl,reprint,superscriptaddress,showpacs]{revtex4-1}
\usepackage{times,graphicx,jab}

\begin{document}
\title{The Nature of Subproton Scale Turbulence in the Solar Wind}
\author{C.~H.~K.~Chen}
\email{chen@ssl.berkeley.edu}
\affiliation{Space Sciences Laboratory, University of California, Berkeley, CA 94720, USA}
\author{S.~Boldyrev}
\author{Q.~Xia}
\affiliation{Department of Physics, University of Wisconsin, Madison, WI 53706, USA}
\author{J.~C.~Perez}
\affiliation{Space Science Center, University of New Hampshire, Durham, NH 03824, USA}
\begin{abstract}
The nature of subproton scale fluctuations in the solar wind is an open question, partly because two similar types of electromagnetic turbulence can occur: kinetic Alfv\'en turbulence and whistler turbulence. These two possibilities, however, have one key qualitative difference: whistler turbulence, unlike kinetic Alfv\'en turbulence, has negligible power in density fluctuations. In this Letter, we present new observational data, as well as analytical and numerical results, to investigate this difference. The results show, for the first time, that the fluctuations well below the proton scale are predominantly kinetic Alfv\'en turbulence, and, if present at all, the whistler fluctuations make up only a small fraction of the total energy.
\end{abstract}
\maketitle

\emph{Introduction}.---Despite many years of observations, the nature of small scale fluctuations in the solar wind remains under debate. In particular, there are several conflicting theories for the turbulence at scales smaller than the proton gyroradius. In this Letter, we present new observations, theory and numerical simulations to determine the types of fluctuations present between the ion and electron scales.

Theoretical descriptions of plasma turbulence can be categorized as weak or strong. Weak turbulence theory involves fluctuations that do not change significantly during each interaction, so that they retain their linear wave mode properties, allowing their energy spectrum to be derived analytically \citep[e.g.,][]{zakharov92,boldyrev95,voitenko98,galtier00,newell01,galtier03a,wang11}. If the turbulence is strong, the non-linear terms in the dynamical equations are comparable to the linear terms and the fluctuations fully decay in one interaction (a situation also known as critical balance \citep{goldreich95}). Therefore, qualitative properties of the linear modes may still be present, even in strong turbulence. Indeed, observations in the solar wind, in which the turbulence is thought to be strong, display many properties that are similar to those of the linear waves \citep[e.g.,][]{belcher71,bale05,he11b,howes12a}.

Knowing the types of fluctuations present is central to understanding the nature of the turbulence. For example, in the solar wind at scales larger than the proton gyroradius, the fluctuations display properties similar to \citet{alfven42} waves \citep[e.g.,][]{belcher71}: perpendicular magnetic fluctuations much larger than parallel magnetic fluctuations ($\delta B_\perp \gg \delta B_\|$), velocity and normalized magnetic fluctuations of similar amplitude $\delta v \sim \delta b$, and frequent times of strong correlation between $\mathbf{v}$ and $\mathbf{b}$. The phase speed of the fluctuations was also suggested to be similar to the Alfv\'en speed \citep{bale05}. These properties are used to justify the application of Alfv\'enic turbulence theory to the solar wind. Interestingly, however, the magnetic fluctuations are slightly larger than the velocity fluctuations $\delta b \gtrsim \delta v$, in both the solar wind and MHD turbulence simulations \citep[][and references therein]{boldyrev11,chen13b}, showing that in strong turbulence there can be quantitative differences to the linear wave relationships.

At smaller scales, around the proton gyroradius and below, the situation is less clear and different wave modes have been suggested to be relevant. Two possibilities are kinetic Alfv\'en waves (KAWs) \citep{leamon98a,leamon99,hollweg99,bale05,howes06,howes08a,howes08b,schekochihin09,chandran09c,shaikh09b,chen10a,howes10,chandran10b,howes11a,howes11c,chandran11,tenbarge12a,boldyrev12b,sahraoui12,mithaiwala12,boldyrev13a,boldyrev13b}, and whistler waves \citep{beinroth81,coroniti82,goldstein94,stawicki01,galtier03a,galtier05,galtier06a,galtier06b,gary08,saito08,cho09a,gary09,shaikh09c,shaikh09b,gary10,saito10,shaikh10b,chang11,gary12,saito12,mithaiwala12}. Since the non-linear equations that they derive from have a similar form, the turbulence energy spectra, obtained from dimensional arguments, are the same.

Previous attempts to distinguish these possibilities in observations considered fluctuations around the proton scale, rather than well below, and led to contradictory or uncertain conclusions. Measurements of the normalized reduced magnetic helicity \citep{goldstein94,leamon98a,hamilton08,he11b,podesta11d,smith12,he12a} indicate that the proton scale fluctuations are generally right handed in the plasma frame, which was initially interpreted as due to the presence of whistler waves \citep{goldstein94}. The KAW, however, is also right handed \citep{gary86,hollweg99,schekochihin09}, so this is not a useful distinguishing measure \citep{howes10}. \citet{bale05} suggested the ratio of electric to magnetic fluctuations at the proton scale to be consistent with the KAW, rather than whistler, dispersion, although \citet{salem12} concluded that this ratio alone was not enough to make the distinction. Various authors have used the amplitude of the parallel, compared to the perpendicular, magnetic fluctuations \citep{gary09,chen10b,smith12,salem12,he12a,sahraoui12,tenbarge12b,podesta12b,kiyani13}, although they reached different conclusions, partly because there is not a large difference between the modes and different definitions of the parallel direction were used \citep{tenbarge12b}. Finally, $k$-filtering, a multi-spacecraft optimization technique, has led to contradictory findings \citep{sahraoui10,narita11,roberts13} and cannot currently be used far below the proton scale, due to the available spacecraft separations.

In this Letter, we present a new measure to clearly distinguish the nature of the fluctuations well below the proton scale. Applying this to solar wind observations, and comparing the result to theory and numerical simulations, shows that the fluctuations between ion and electron scales are predominantly kinetic Alfv\'en turbulence, rather than whistler turbulence.

\emph{Theory}.---In a collisionless plasma of beta $\beta\sim 1$, both the KAW and whistler wave can be excited for perpendicular scales between the ion and electron gyroradii, $1/\rho_i\ll k_\perp\ll 1/\rho_e$, \citep[e.g.,][]{boldyrev13a}. While some properties of these electromagnetic modes are qualitatively similar, such as the dispersion relation, magnetic compressibility and helicity, there is one key difference. The KAW is low frequency compared to the ion thermal speed, $\omega\ll k_\perp v_{th,i}$, so the ions can fluctuate and are involved in the dynamics, along with the electrons. The whistler wave, however, is high frequency,  $\omega\gg k_\perp v_{th,i}$, so the ions are not able to move fast enough and, due to quasi-neutrality, the electron density fluctuations are also negligibly small. This can be seen, for example, in the numerical solutions of \citet{gary09} and the analytical solutions of \citet{boldyrev13a}. 

It is possible, however, for whistler waves to generate density fluctuations if their frequency is not asymptotically large, but close to $k_\perp v_{th,i}$ so that the ions can still fluctuate. To produce density fluctuations, $\delta n/n_0$, comparable to the magnetic fluctuations, $\delta B/B_0$, would require in this case all of the energy to be at $k_\|d_i\approx 1$ (where $d_i$ is the ion inertial length) for any $k_\perp$. A slight spread of energy over  $k_\|$ would significantly decrease the density amplitude according to $(\delta n/n_0)^2/(\delta B/B_0)^2 \sim \frac{1}{2} (k_\|d_i)^{-4}$ \citep{boldyrev13a}. However, for $k_\|d_i\approx 1$ the whistler modes would be strongly damped by the ions: the damping rate is $\gamma/\omega_0=-2\sqrt{\pi}\beta_i^{-1}x\exp(-x^2)$, where $x=k_\|d_i/\sqrt{\beta_i}$ \citep{boldyrev13a}, so for them to remain undamped at $\beta_i\sim 1$ would require $x\gtrsim2$. For such values of $x$ these modes would have very small density fluctuations: $(\delta n/n_0)^2/(\delta B/B_0)^2 \lesssim 0.03$. This difference between the density fluctuations in whistlers and KAWs is the basis of our technique to distinguish the nature of the fluctuations.

A natural normalization for the density and magnetic fluctuations in kinetic Alfv\'en turbulence is \citep{schekochihin09,boldyrev13b}
\begin{eqnarray}
\label{eq:nnorm}
\delta\tilde{n}&=&\left(1+\frac{T_i}{T_e}\right)^\frac{1}{2}\frac{v_s}{v_A}\left[1+\left(\frac{v_s}{v_A}\right)^2\left(1+\frac{T_i}{T_e}\right)\right]^\frac{1}{2}\frac{\delta n}{n_0},\\
\label{eq:bnorm}
\delta \tilde{\mathbf{b}}&=&\frac{\delta\mathbf{B}}{B_0}, 
\end{eqnarray}
where $T_i$ and $T_e$ are the ion and electron temperatures, $v_s=\sqrt{T_e/m_i}$ is the ion acoustic speed, $m_i$ is the ion mass, $v_A$ is the Alfv\'en speed, $n_0$ is the mean density and $B_0$ is the mean magnetic field strength. With this normalization, the KAW has equal density and perpendicular magnetic fluctuation amplitudes, $\delta\tilde{b}_\perp=\delta\tilde{n}$, independent of the wavevector. For the whistler wave, $\delta\tilde{b}_\perp\gg\delta\tilde{n}$ for the reasons discussed above. We note that these equations do not include temperature anisotropies and the small population of alpha particles which drifts with respect to the protons in the solar wind. While these features can affect such normalizations \citep{chen13b}, they are not significant, compared to the error estimates, for the data intervals considered here.

\emph{Results}.---The \emph{ARTEMIS} spacecraft \citep{angelopoulos11} have measured the density and magnetic fluctuations in the solar wind at 1 AU with sufficient resolution to test these predictions. The spacecraft potential fluctuations measured by EFI \citep{bonnell08} were used to infer the electron density fluctuations \citep{chen12a,chen13a} and the magnetic fluctuations were measured by SCM \citep{roux08}. For the normalization factors and kinetic scales, FGM \citep{auster08} was used for the DC magnetic field and the ESA ground moments \citep{mcfadden08a} were used for the particle densities, bulk velocities and temperatures.

\begin{figure}
\includegraphics[scale=0.45]{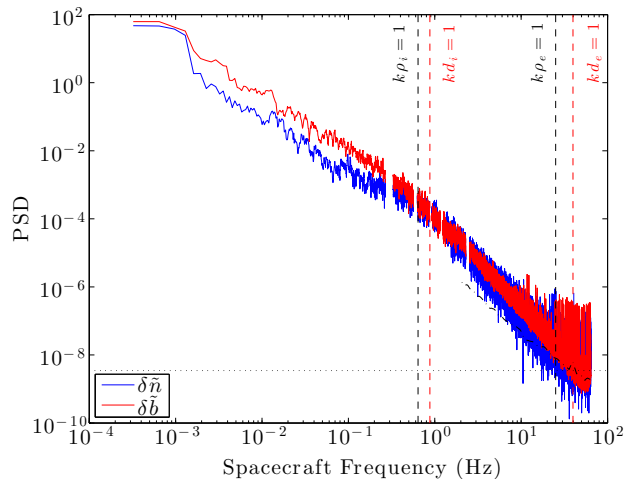}
\caption{\label{fig:spectra}Spectrum of density and magnetic fluctuations in the solar wind, normalized according to Equations \ref{eq:nnorm} and \ref{eq:bnorm}. Between ion and electron scales (vertical dashed lines) the spectra are of similar amplitude, which suggests the presence of kinetic Alfv\'en turbulence, rather than whistler turbulence.}
\end{figure}

Figure \ref{fig:spectra} shows the spectra of electron density and magnetic fluctuations, normalized according to Equations \ref{eq:nnorm} and \ref{eq:bnorm}, measured by \emph{ARTEMIS-P2} on 11th October 2010 from 00:21 to 01:14 UT. They have a similar shape to such kinetic scale spectra measured by other spacecraft \citep[e.g.,][]{chen10b,safrankova13a}. The magnetic field spectrum uses FGM data below 2 Hz and SCM data above 2 Hz. Noise due to harmonics of the spacecraft spin frequency (0.30 Hz), clock frequencies (8/32 Hz) and sidebands due to the spin modulation of the clock frequencies were removed by deleting the affected portions of the spectrum. There is additional noise at higher spacecraft-frame frequencies ($f_{sc}>$ 10 Hz), but this is outside our range of current interest. The density fluctuation noise floor is marked with a horizontal dotted line and the SCM noise floor with a dash-dotted line. Proton and electron gyroradii, $\rho_{i,e}$, and inertial lengths, $d_{i,e}$, are marked assuming the \citet{taylor38} hypothesis. It can be seen that between the ion and electron scales, the normalized density and magnetic fluctuations are of similar amplitude, $\delta\tilde{b}\sim\delta\tilde{n}$. As discussed above, this suggests that the turbulence is kinetic Alfv\'en in nature, rather than whistler.

\begin{figure}
\includegraphics[scale=0.45]{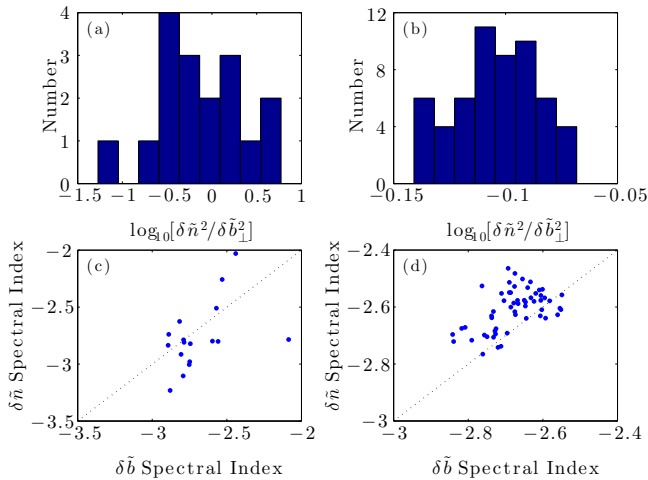}
\caption{\label{fig:comparison}Histograms of the kinetic Alfv\'en ratio in (a) the solar wind and (b) kinetic Alfv\'en turbulence simulation. Scatterplots of spectral indices in (c) the solar wind and (d) kinetic Alfv\'en turbulence simulation.}
\end{figure}

The same analysis was performed on all of the the 17 intervals used by \citet{chen12a,chen13a}. This number is limited since the spacecraft need to be in the free solar wind \citep{chen11b} and in burst mode. The proton beta for these intervals covers the range $0.29\leq\beta_i\leq 3.7$, and the electron beta $0.40\leq\beta_e\leq 5.5$. For each interval, the average ``kinetic Alfv\'en ratio'', defined as $\delta\tilde{n}^2/\delta\tilde{b}_\perp^2$, was calculated over the range $2.5<f_{sc}<7.5$ Hz. This range was chosen because it is high enough to avoid spacecraft spin effects in the magnetic field spectrum, low enough to avoid the instrument noise floors and is between the proton and electron scales $5\lesssim k\rho_i\lesssim14$. Since the total magnetic energy, $\delta\tilde{b}^2=\delta\tilde{b}_\perp^2+\delta\tilde{b}_\|^2$, was measured with the SCM data, the perpendicular fluctuations were obtained using
\begin{equation}
\label{eq:btobperp}
\delta\tilde{b}_\perp^2=\delta\tilde{b}^2-\frac{\left(v_s^2/v_A^2\right)\left(1+T_i/T_e\right)}{1+\left(v_s^2/v_A^2\right)\left(1+T_i/T_e\right)}\delta\tilde{n}^2.
\end{equation}
This relationship comes from the non-linear kinetic Alfv\'en equations \citep{schekochihin09,boldyrev13a} and avoids the need to define the parallel direction. The distribution of the average kinetic Alfv\'en ratio in the 17 solar wind intervals is shown in Figure \ref{fig:comparison}a. The geometric mean is $\delta\tilde{n}^2/\delta\tilde{b}_\perp^2=0.75$, indicating approximate equipartition with a slight excess of magnetic energy. 

To compare the solar wind measurements with expectations for strong kinetic Alfv\'en turbulence, a similar analysis was performed on a numerical simulation of the non-linear kinetic Alfv\'en equations \citep{boldyrev13a}. The equations were simulated in a 512$^3$ periodic box, forced in the strong regime and are described in more detail elsewhere \citep{boldyrev12b,boldyrev13a,boldyrev13b}. The distribution of the kinetic Alfv\'en ratio in 56 snapshots is shown in Figure \ref{fig:comparison}b. The geometric mean is $\delta\tilde{n}^2/\delta\tilde{b}_\perp^2=0.79$, similar to the solar wind value. This similarity is good evidence for the kinetic Alfv\'en nature of solar wind turbulence between the ion and electron scales.

One difference between the solar wind and the simulation is the spread of values of the kinetic Alfv\'en ratio: the standard deviation of the solar wind values is 26 times larger. This spread, however, is not expected to be universal and may depend on several factors. Firstly, there are many more data points in each simulation snapshot than in each solar wind interval: the average interval length is 13 min, which is $\sim10^5$ data points, compared to $\sim10^8$ in a simulation snapshot. Secondly, there are significant uncertainties in the measured quantities used in the normalization factors (Equations \ref{eq:nnorm} and \ref{eq:bnorm}) that are not present in the simulation. Thirdly, the possibility of some whistler turbulence causing some of the spread in the data (towards lower kinetic Alfv\'en ratios) cannot be ruled out, although the energy in such fluctuations cannot generally be more than a few per cent. Finally, there may be other physics occurring in the solar wind at kinetic scales, such as instabilities, that is not captured in the simulation. The mean values of the kinetic Alfv\'en ratio, along with their standard error of the mean are given in Table \ref{tab:results}.

The final comparison performed between the solar wind and the simulation is the correlation between the spectral indices of the different fields. Scatterplots of the spectral indices measured in the solar wind intervals over the range $2.5<f_{sc}<7.5$ Hz and in the simulation snapshots taken from \cite{boldyrev12b} with fitting range $5<k<15$, are shown in Figure \ref{fig:comparison}c,d. Again, the spread in values is smaller in the simulation, as expected, but the linear correlation coefficients $r$ (given in Table \ref{tab:results}, along with the 95\% confidence intervals) are similar: there is a mild positive correlation. This similarity is further evidence that the kinetic scale fluctuations in the solar wind can be described by kinetic Alfv\'en turbulence.

\emph{Discussion}.---We have shown that between ion and electron scales in the solar wind, the kinetic Alfv\'en ratio is $\delta\tilde{n}^2/\delta\tilde{b}_\perp^2=0.75$, as expected for kinetic Alfv\'en turbulence, rather than whistler turbulence, in which the ratio should be smaller by more than an order of magnitude. Although the density and perpendicular magnetic fluctuations are of similar amplitude, there is a slight excess of magnetic energy in both the solar wind observations and the kinetic Alfv\'en turbulence simulation. This suggests that strong kinetic Alfv\'en turbulence, while having some properties of KAWs, can produce quantitative differences, similarly to the excess magnetic energy seen at MHD scales \citep[e.g.,][]{boldyrev11,chen13b}.

\begin{table}
\caption{\label{tab:results}Comparison of solar wind data and kinetic Alfv\'en turbulence simulation}
\begin{ruledtabular}
\begin{tabular}{cccc}
& $\delta\tilde{n}^2/\delta\tilde{b}_\perp^2$ & $r$ \\
\hline
solar wind & $0.75^{+0.22}_{-0.17}$ & $0.46^{+0.31}_{-0.49}$ \\
simulation & $0.786^{+0.004}_{-0.004}$ & $0.52^{+0.17}_{-0.22}$ \\
\end{tabular}
\end{ruledtabular}
\end{table}

In the estimates of whistler compressibility discussed earlier, the results of linear theory were used. Due to the critical balance condition, one may expect this estimate to approximately hold for strong turbulence as well. However, such an assumption may not be necessary as there is a reason to believe that if whistler turbulence is present in this case, it would be weak. Indeed, at $k_\perp\rho_i=7.1$ for the interval in Figure \ref{fig:spectra}, the fluctuation amplitude is $\delta B/B_0=0.026$. Strong turbulence requires $\delta B/B_0=k_\|/k_\perp$, giving $k_\| \rho_i=0.18$ and a propagation angle of $\theta=88.5^\circ$. For whistlers to exist at such $k_\perp$ requires $k_\|/k_\perp>\sqrt{2\beta_i}/(k_\perp\rho_i)$, that is, $\theta<75^\circ$, since $\beta_i=1.96$ here. Since the required angles are larger, this means that if the turbulence is strong, it cannot be whistler turbulence. Having $\theta=75^\circ$ would mean $\delta B/B_0\ll k_\|/k_\perp$ so whistler turbulence would be in the weak regime. The fact that the fluctuations in the solar wind are strongly non-Gaussian in this range \citep{alexandrova08b,kiyani09a,kiyani13} lends further support to the strong kinetic Alfv\'en turbulence interpretation. 

Since solar wind turbulence at MHD scales is predominantly Alfv\'enic, with around 10\% of the energy in the slow mode fluctuations and very little in the fast mode, it makes sense that the transition is to kinetic Alfv\'en, rather than whistler, turbulence \citep{howes12a}. This is also consistent with other observations, such as the fluctuations being anisotropic with $k_\perp>k_\|$ \citep{chen10b}, having a significant, rather than negligible, parallel electric field spectrum \citep{mozer13}, and the flattening of the density spectra at ion scales \citep{chen13a}, which is thought to be due to the enhanced compressibility of kinetic Alfv\'en turbulence \citep{chandran09c}. Since such fluctuations are relatively low frequency, the Taylor hypothesis can be used to relate the spacecraft-frame frequency spectra to the wavenumber spectra of theory and simulations. Indeed, the measured spectral indices of density and magnetic fluctuations $\approx$ --2.7 are similar to those in kinetic Alfv\'en turbulence simulations \citep{howes11a,boldyrev12b}.

If the frequency of the subproton scale fluctuations becomes close to the ion cyclotron frequency $\omega -\Omega_i\sim \Omega_i/\sqrt{k_\perp \rho_i}$, it has been proposed that ion-Bernstein modes may couple to the turbulent cascade \citep[][]{howes08a,howes09,podesta12a}. For a kinetic Alfv\'en cascade, this would happen when $k_\|v_{th, i}\sim \Omega_i/(k_\perp \rho_i)$. For collisionless damping of these modes to be negligible, the ratio $k_\|v_{th, i}/(\omega -\Omega_i)\sim 1/\sqrt{k_\perp \rho_i}$ should be much smaller than one. It then follows that in the asymptotic limit $k_\perp \rho_i \gg 1$ the ion-Bernstein modes occupy a narrow band in the frequency space, which may reduce their coupling to the kinetic Alfv\'en cascade \cite[e.g.,][]{howes08a,howes09}. For moderate values of $k_\perp \rho_i$, on  the other hand, these modes are relatively strongly damped compared to the kinetic Alfv\'en modes \cite[][]{podesta12a}, which also reduces their energetic relevance. A quantitative estimate of the level of ion-Bernstein fluctuations driven by  kinetic Alfv\'en turbulence must await further observations and fully kinetic numerical treatment.

The nature of the subproton scale fluctuations has some important implications, such as understanding plasma heating. For example, since the fluctuations are kinetic Alfv\'en turbulence, the cyclotron resonance may not be as important relative to other damping mechanisms \citep{schekochihin09}. Kinetic Alfv\'en turbulence may also generate particular types of structures, such as 2D sheets \citep{boldyrev12b}, which are important for understanding heating if it occurs preferentially at such structures \citep{osman12a,tenbarge13}. Determining the nature of subproton scale turbulence is also relevant to other astrophysical plasmas, such as the ionized interstellar medium \citep{armstrong95} and hot accretion flows \citep{quataert98}, which may have similar turbulence properties but are not as well measured. Understanding subproton scale turbulence in the solar wind, such as the fact that it is predominantly kinetic Alfv\'en in nature, can provide insight into the dynamics and heating of such plasmas.

\begin{acknowledgments}
This work was supported by NASA contracts NNN06AA01C and NAS5-02099, US DoE awards DE-FG02-07ER54932, DE-SC0003888 and DE-SC0001794, NSF grant PHY-0903872, and the NSF Center for Magnetic Self-Organization in Laboratory and Astrophysical Plasmas at U. Wisconsin-Madison. We acknowledge the \emph{THEMIS}/\emph{ARTEMIS} team and thank O. Le Contel for useful discussions regarding the SCM instrument. High Performance Computing resources were provided by the Texas Advanced Computing Center (TACC) at the University of Texas at Austin under the NSF-Teragrid Project TG-PHY120042.
\end{acknowledgments}

\bibliography{bibliography}

\end{document}